# Modélisation et analyse de l'erreur statique de transmission d'un engrenage. Influence des déformations des roues et interactions entre les couples de dents en prise

# Modelling and Analysis of Static Transmission Error of Gears. Effect of Wheel Body Deformation and Interactions between Adjacent Loaded Teeth

Emmanuel RIGAUD* et Denis BARDAY**

* Laboratoire de Tribologie et Dynamique des Systèmes. UMR 5513. Ecole Centrale de Lyon.
36, avenue Guy de Collongue. B.P. 163. 69131 Ecully Cedex.

** Direction des Etudes et Recherches. RENAULT V. I.
1, Avenue H. Germain. B.P. 310. 69802 Saint-Priest Cedex.

**Résumé :** L'excitation vibratoire des transmissions résulte des écarts de géométrie et des déformations élastiques des engrenages. Pour calculer ses caractéristiques, nous avons introduit une modélisation par éléments finis 3D des roues dentées puis résolu le système d'équations non linéaires qui gère le contact entre les dentures. A partir de cette approche, nous avons pu mettre en évidence les interactions entre les couples de dents en prise et les couplages élastiques entre les corps de roue et les dentures. Nous avons aussi analysé les effets des corrections de denture sur les fluctuations périodiques de l'erreur statique de transmission sous charge et de la raideur d'engrènement.

**Abstract :** We have developped a numerical tool in order to predict vibratory excitation of a gearbox induced by the meshing process. We propose a method based on a 3D Finite Element modelling of each toothed wheel. A non linear algorithm computes the fluctuations of static transmission error and mesh stiffness. Numerical simulations show the effect of wheel body deformation and the interactions between adjacent loaded teeth. The method developped allows to analyse the effect of different tooth modification types an manufacturing errors on the static transmission error.

## 1. INTRODUCTION

La principale source d'excitation vibratoire des transmissions par engrenages est générée par le processus d'engrènement et ses caractéristiques dépendent des situations instantanées des couples de dents en prise. Ces situations résultent essentiellement des écarts de géométrie et des déformations élastostatiques des engrenages. On suppose généralement que c'est l'erreur statique de transmission sous charge qui constitue la principale source d'excitation interne d'une transmission [1]. Cette erreur correspond, pour une vitesse de rotation très faible et sous l'application du couple moteur, à l'écart entre la position réelle occupée par la roue menée et sa position théorique. En régime de fonctionnement stationnaire, elle est à l'origine d'une fluctuation périodique de la raideur d'engrènement et d'une excitation de type déplacement qui génèrent des surcharges dynamiques sur les dentures. Celles-ci sont transmises aux lignes d'arbres, aux roulements et au carter. Les vibrations qui résultent de l'excitation de ces différentes composantes sont à l'origine de nuisances acoustiques.

Nous présentons ci-après un outil numérique qui permet de calculer l'évolution temporelle de l'erreur statique de transmission sous charge d'un engrenage à denture droite ou hélicoïdale, à partir de ses caractéristiques géométriques, de ses corrections de forme, de ses défauts et des conditions de fonctionnement (couple moteur). Cet outil s'appuie sur les calculs des déformations en flexion des dents d'engrenages et des déformations locales de contact. La résolution des équations non linéaires qui gèrent le contact entre les roues dentées permet de calculer, pour différentes positions successives de la roue menante, le rapprochement des dents en prise.

## 2. METHODE DE CALCUL ADOPTEE

### 2.1 Calcul des déformations élastostatiques des engrenages

Le calcul de l'erreur statique de transmission sous charge nécessite en premier lieu l'estimation des déformations élastostatiques des dents en prise. Différents auteurs ont calculé celles-ci à partir de modèles de poutres [2], de plaques d'épaisseur variable [3], ou bien à partir de discrétisation des dents en 2 dimensions [4]. Les hypothèses associées à ce type de modèles sont difficilement justifiables car les dimensions caractéristiques des dents ne sont représentatives ni d'une poutre, ni d'une plaque et le comportement d'une dent évolue selon la largeur. Bien que ces hypothèses n'aient pas été vérifiées, les différents auteurs qui exploitent des modèles éléments finis en 3 dimensions [5] supposent généralement que les déformations des dents sont découplées les unes des autres et que la distribution de la charge sur les dents en prise n'est pas affectée par les déformations des corps des roues dentées.

Ces observations nous ont conduit à développer un modèle éléments finis 3D de chaque roue dentée original. Nous avons décrit précisément la géométrie de l'engrenage (profil en développante de cercle, trochoïde de raccordement en pied de dent) en raison de l'influence sensible des paramètres de conception sur la déformation en flexion des dents. Pour chaque roue dentée, nous avons pris en considération plusieurs dents successives ainsi que l'ensemble du corps de la roue. Nous avons utilisé des éléments finis de forme hexaédrique à 20 noeuds et 3 degrés de liberté par noeuds. Nous avons supposé que la roue dentée était encastrée sur son arbre. Le maillage de chaque roue dentée possède environ 8000 noeuds, 1500 éléments et 17000 degrés de liberté. A titre d'illustration, la figure 1 présente le maillage d'un engrenage parallèle à denture hélicoïdale 35/49 dents. A partir de ce modèle, nous effectuons un certain nombre de calculs statiques qui permettent de calculer la matrice de souplesse $\underline{\underline{\mathbf{H}}}^{\mathbf{u,F}}\mathbf{(\omega=0)}$ associée aux noeuds des surfaces des dents et définie comme suit : $\underline{\mathbf{u}} = \underline{\underline{\mathbf{H}}}^{\mathbf{u,F}}(\omega = 0).\underline{\mathbf{F}}$.

## 2.2 Calcul de l'erreur statique de transmission sous charge et de la raideur d'engrènement

On définit l'erreur statique de transmission sous charge comme le rapprochement des dents $\delta$ exprimé sur la ligne d'action de l'engrenage. Pour évaluer son évolution temporelle, on calcule $\delta$ pour plusieurs positions successives $\theta$ de la roue menante.

Pour chaque position $\theta$, une analyse cinématique du fonctionnement de l'engrenage nous permet de déterminer la ligne de contact (lieu des contacts sur les surfaces des dents en prise). Une fois cette ligne de contact discrétisée, le système d'équations qui gère les déformations élastostatiques de l'engrenage peut s'écrire :

$$\begin{cases} \underline{\underline{\mathbf{H}}}^{u,F}(\omega=0).\underline{\mathbf{F}} = \underline{\delta}(\theta) - \underline{\mathbf{e}} - \underline{\mathbf{hertz}}(\underline{\mathbf{F}}) \\ \Sigma\, \mathbf{F[i]} = \mathbf{F}_{total} \end{cases} \quad \mathbf{(1)}$$

$\underline{\underline{\mathbf{H}}}^{u,F}(\omega=0)$ est la matrice de souplesse associée aux noeuds de la ligne de contact et calculée à partir des matrices de souplesse de chaque roue dentée; $\underline{\delta}(\theta)$ est un vecteur dont les composantes, toutes identiques, correspondent au rapprochement des dents; $\underline{\mathbf{F}}$ correspond au vecteur des efforts s'exerçant sur les dents; $\underline{\mathbf{e}}$ correspond au vecteur des écarts initiaux séparant les dents et induits par les défauts de géométrie et les corrections de denture; $\underline{\mathbf{hertz}}$ correspond au vecteur des déplacements induits par les déformations locales de contact et calculés à partir de la théorie de Hertz; $\mathbf{F}_{total}$ correspond à l'effort total s'exerçant sur l'ensemble des couples de dents en prise et induit par le couple moteur.

Les inconnues du système (1) composé de (N+1) équations sont les N composantes du vecteur $\underline{\mathbf{F}}$ qui s'appliquent sur les N noeuds de la ligne de contact et la valeur du rapprochement $\delta(\theta)$. Ce système est non linéaire pour deux raisons. D'abord, chaque dent entre en contact de façon progressive en fonction de la charge, ce qui induit une évolution de la ligne de contact réelle (non-linéarité géométrique). Ensuite, les déformations locales des contacts hertziens sont non linéaires. L'ordre de grandeur de ces déformations correspond à celui des déformations en flexion des dents d'engrenage et elles contribuent de façon significative à l'erreur statique de transmission sous charge.

Nous résolvons le système (1) à l'aide d'un processus itératif qui nous permet de calculer la répartition finale de la charge et la valeur de l'erreur statique de transmission $\delta(\theta)$.

Pour chaque position de la roue menante, la raideur d'engrènement est définie par une linéarisation, autour de la position d'équilibre statique, de la courbe qui traduit l'évolution de l'erreur statique de transmission sous charge en fonction de la charge appliquée $F_s$ :

$$\mathbf{K}(\mathbf{F}_s,\theta) = \frac{\partial(\mathbf{F}_s)}{\partial(\delta(\mathbf{F}_s,\theta))}$$

Pour un engrenage sans défauts de géométrie, les évolutions temporelles de l'erreur statique de transmission et de la raideur d'engrènement sont périodiques, de période fondamentale égale à la période d'engrènement.

## 3. RESULTATS NUMERIQUES

Les résultats qui suivent correspondent à l'engrenage parallèle 35/49 dents à denture hélicoïdale présenté sur la figure 1. Ses caractéristiques géométriques correspondent à celles d'un engrenage équipant une boîte de vitesses de camion (module de 3,5 mm et largeur de 36,5 mm).

### 3.1 Influence de l'élasticité des roues

Pour mettre en évidence l'influence de l'élasticité des corps de roue, nous avons considéré trois modélisations distinctes de l'engrenage. Dans la première configuration, les corps de roue 35 et 49 dents sont rigides (seule la jante de la roue est élastique). La seconde configuration correspond à deux roues élastiques pleines. La troisième intègre le voile mince de la roue 49 dents.

La prise en compte de l'élasticité des roues modifie considérablement l'ordre de grandeur et la répartition des déformations sur les surfaces actives des dents. La figure 2 montre que la valeur moyenne de l'erreur statique de transmission sous charge augmente de façon significative (31 µm pour des corps de roue rigides, 39 µm pour des roues élastiques pleines et 67 µm si on prend en compte le voile de la roue 49 dents). Les couplages élastiques entre les corps de roue et la denture entraînent aussi une chute de la valeur moyenne de la raideur d'engrènement.

Les efforts s'exerçant sur la ligne de contact sont à l'origine d'un moment qui induit une déformation en flexion de la roue et qui entraîne un déplacement angulaire important des dents en prise. L'amplitude de ce moment dépend de la répartition de la charge sur la ligne de contact et, par conséquent, de la position de la roue menante. L'élasticité des roues est donc à l'origine d'une modification des conditions de contact entre les dents en prise. Elle entraîne une modification de la forme des fluctuations de l'erreur statique de transmission sous charge et de son amplitude crête à crête. Celle-ci est égale à 2.7 µm pour des corps de roue rigides, à 3.1 µm pour des roues élastiques pleines et à 6.3 µm si on prend en compte le voile de la roue 49 dents.

### 3.2 Interactions entre les couples de dents en prise

Nous considérons le cas de l'engrenage composé d'une roue 35 dents pleine et d'une roue 49 dents présentant un voile mince.

Le chargement appliqué sur un noeud entraîne des déformations sur toute la surface active de la dent chargée ainsi que des déformations des dents précédente et suivante. Ces dernières contribuent à l'erreur statique de transmission sous charge car, le rapport de conduite étant supérieur à 1, plusieurs couples de dents participent simultanément à l'engrènement.

Nous avons calculé les fluctuations de l'erreur statique de transmission sous charge et de la raideur d'engrènement en éliminant les coefficients $\mathbf{H^{u,F}(I, J; \omega=0)}$ qui traduisent l'effet de l'effort s'exerçant sur une dent sur les déformations des autres couples de dents présents dans la zone de contact.

La figure 3 montre qu'il est nécessaire de prendre en compte les interactions entre les différents couples de dents en prise pour prédire correctement les caractéristiques de l'excitation vibratoire. En effet, ces interactions modifient sensiblement aussi bien les valeurs moyennes que les amplitudes crête à crête de l'erreur statique de transmission sous charge et de la raideur d'engrènement. Bien que cette technique soit largement utilisée, il n'est donc pas possible de calculer l'excitation vibratoire à partir de l'estimation de la raideur d'un seul couple de dents.

### 3.3 Introduction des corrections de denture

Les engrenages sont généralement corrigés. On modifie volontairement la forme des dents pour permettre de limiter les chocs au moment de l'entrée en contact des dents, pour optimiser la répartition de la charge le long de la ligne de contact et pour minimiser les fluctuations de l'erreur statique de transmission sous charge.

Grâce à l'outil numérique développé, nous avons pu rechercher les corrections optimales qui permettent de diminuer, pour un couple moteur de 1300 Nm, les fluctuations de l'erreur statique de transmission sous charge (figure 4). L'amplitude crête à crête de celle-ci est égale à 1.5 µm contre 6.5 µm pour la denture non corrigée.

Les corrections introduites correspondent à des enlèvements de matière dans la direction du profil ou dans la direction longitudinale. Les simulations numériques que nous avons effectuées montrent qu'elles ne modifient que très peu l'évolution temporelle de la raideur d'engrènement. En effet, lorsque le couple de charge est suffisant pour que le contact s'établisse sur toute la longueur totale de contact potentielle, la non-linéarité de l'erreur statique de transmission n'est plus induite que par la (faible) non-linéarité des déformations hertziennes. La raideur d'engrènement devient alors presque indépendante de la charge.

En définitive, les corrections qui permettent de minimiser l'amplitude crête à crête de l'erreur statique de transmission sous charge ne permettent pas de minimiser les fluctuations de la raideur d'engrènement. Certains auteurs qui définissent la raideur comme le rapport entre la charge appliquée et le rapprochement des dents en prise, aboutissent donc à des conclusions contraires. Selon eux, une optimisation des corrections de denture qui conduit à une réduction des fluctuations de l'erreur de transmission entraîne forcément une réduction simultanée des fluctuations de la raideur d'engrènement. La figure 4b montre par ailleurs que cette définition de la raideur d'engrènement conduit à une estimation erronée de sa valeur moyenne (220 N/µm contre 320 N/µm).

## 4. CONCLUSION

Nous avons développé un outil numérique pour calculer les fluctuations de l'erreur statique de transmission sous charge et de la raideur d'engrènement d'engrenages parallèles à denture droite ou hélicoïdale, présentant éventuellement des voiles minces et des écarts de forme. Les nombreuses simulations numériques effectuées nous ont permis de mettre en évidence un certain nombre de phénomènes physiques qui sont en partie ignorés par les autres modèles de calcul. Tout d'abord, il est nécessaire de prendre en compte la géométrie exacte de l'engrenage pour bien prédire ses déformations. Ensuite, les interactions entre les différents couples de dents en prise et les couplages entre les corps de roue et les dentures augmentent considérablement la valeur moyenne de l'erreur statique de transmission sous charge et diminuent celle de la raideur d'engrènement. Ces couplages élastiques modifient la forme des fluctuations et l'amplitude crête à crête de ces grandeurs. Ces différentes remarques justifient les choix que nous avons fait lors du développement du modèle éléments finis 3D qui nous a permis de calculer les déformations élastiques de chaque roue dentée.

Nous avons analysé l'effet des corrections de denture. Les corrections qui permettent de minimiser l'amplitude crête à crête de l'erreur statique de transmission sous charge ne permettent pas de minimiser les fluctuations de la raideur d'engrènement. Les corrections optimales évoluent avec le niveau de l'erreur statique de transmission sous charge. Elles dépendent donc du couple moteur transmis, de la géométrie des roues, mais aussi des méthodes employées et des hypothèses retenues pour calculer l'erreur statique de transmission sous charge.

En définitive, la détermination de l'erreur statique de transmission et de la raideur d'engrènement et la recherche des corrections qui minimisent les fluctuations de ces grandeurs doivent s'appuyer sur des méthodes qui intègrent l'ensemble des phénomènes physiques qui contribuent aux déformations élastostatiques des engrenages.

**BIBLIOGRAPHIE**

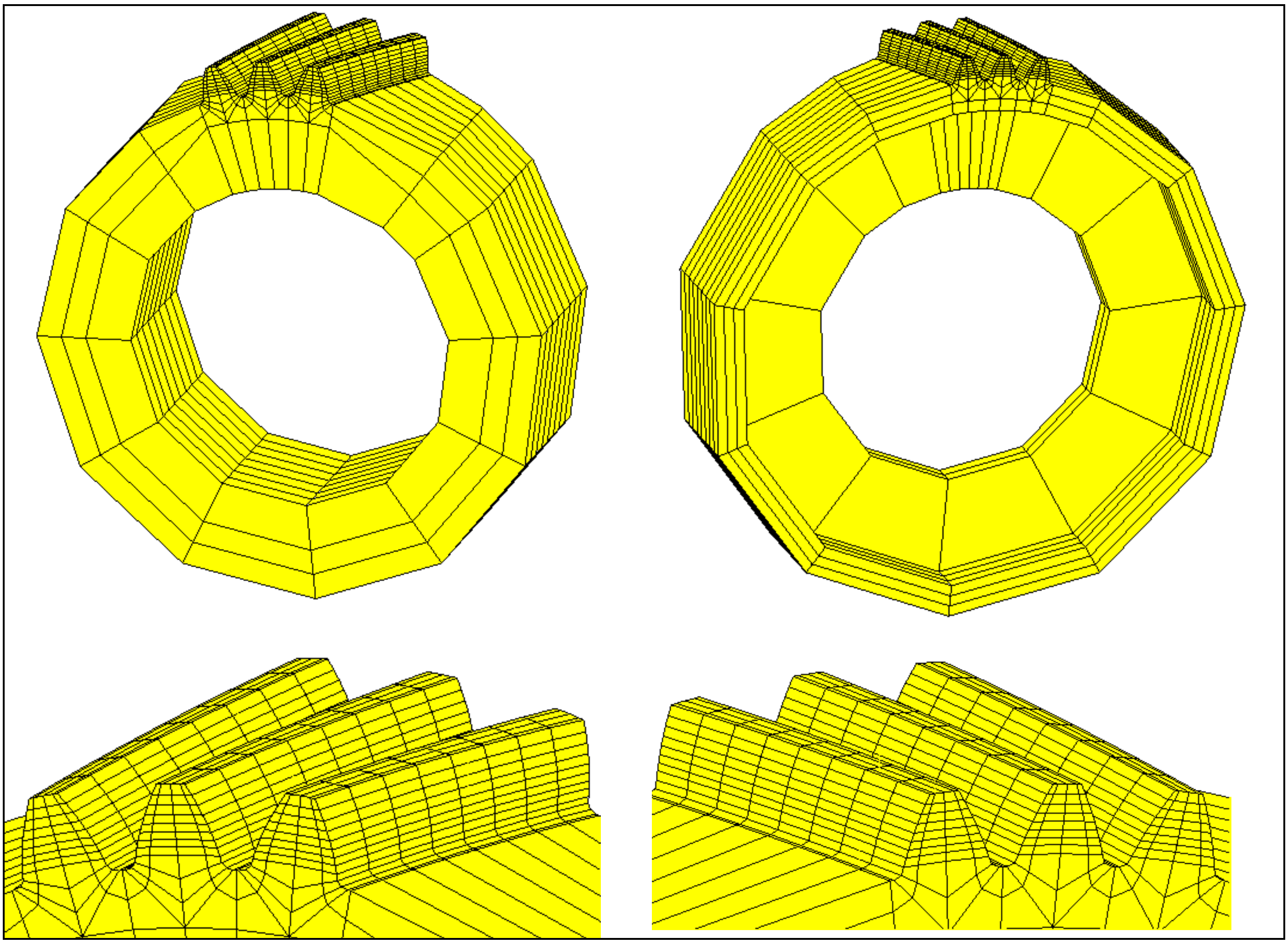

Figure 1. Maillage des roues dentées.

*Figure 1. Meshing of the toothed wheels.*

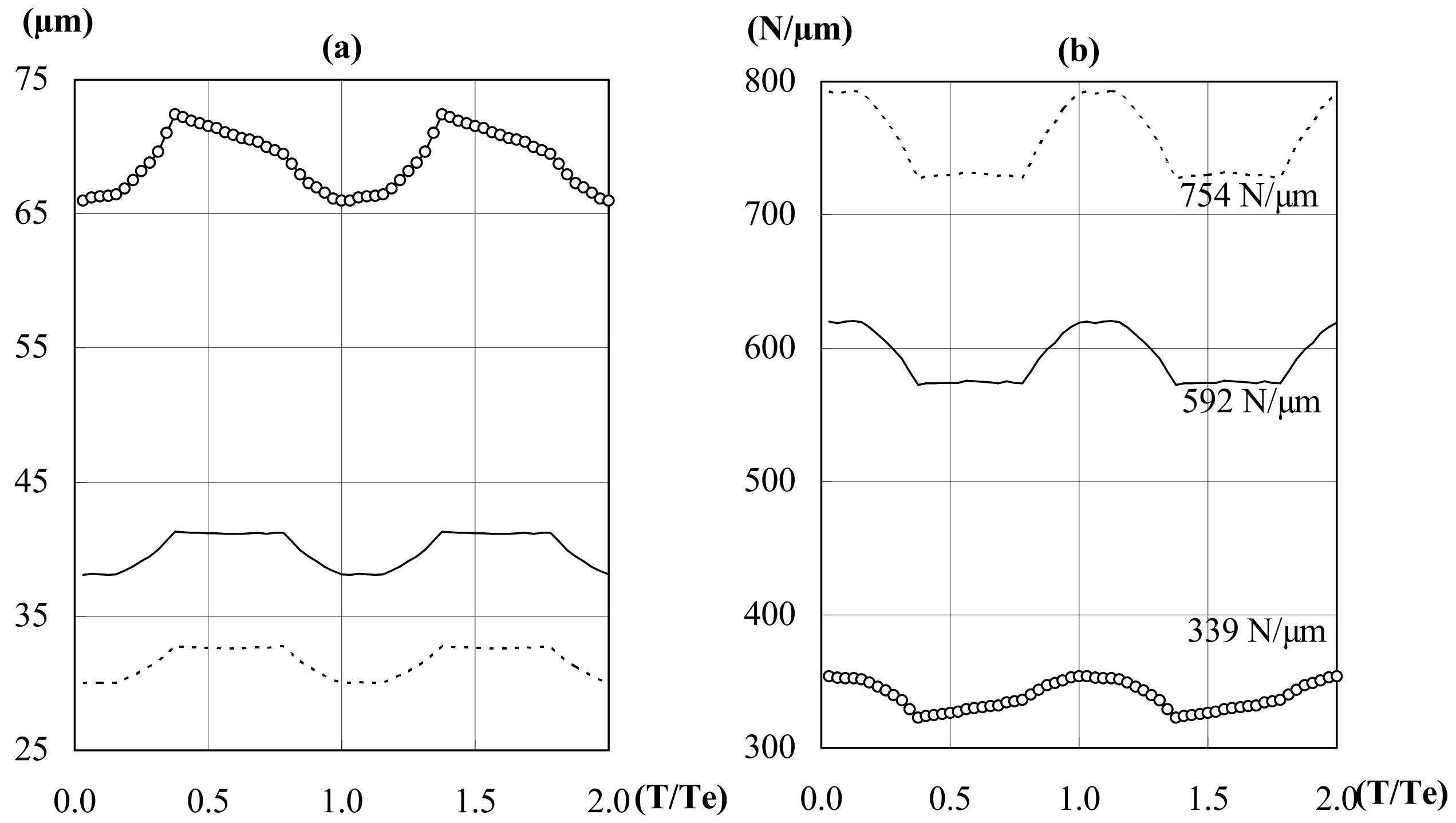

Figure 2. Evolutions temporelles de l'erreur statique de transmission sous charge **(a)** et de la raideur d'engrènement **(b)**.

Configuration réelle (oooooooo), roues élastiques pleines (———), corps des roues rigides (--------).

*Figure 2. Static transmission error* **(a)** *and mesh stiffness* **(b).**

*solid elastic pinion and thin-rimmed wheel* (oooooooo), *solid elastic wheel bodies* (———), *rigid wheel bodies* (--------).

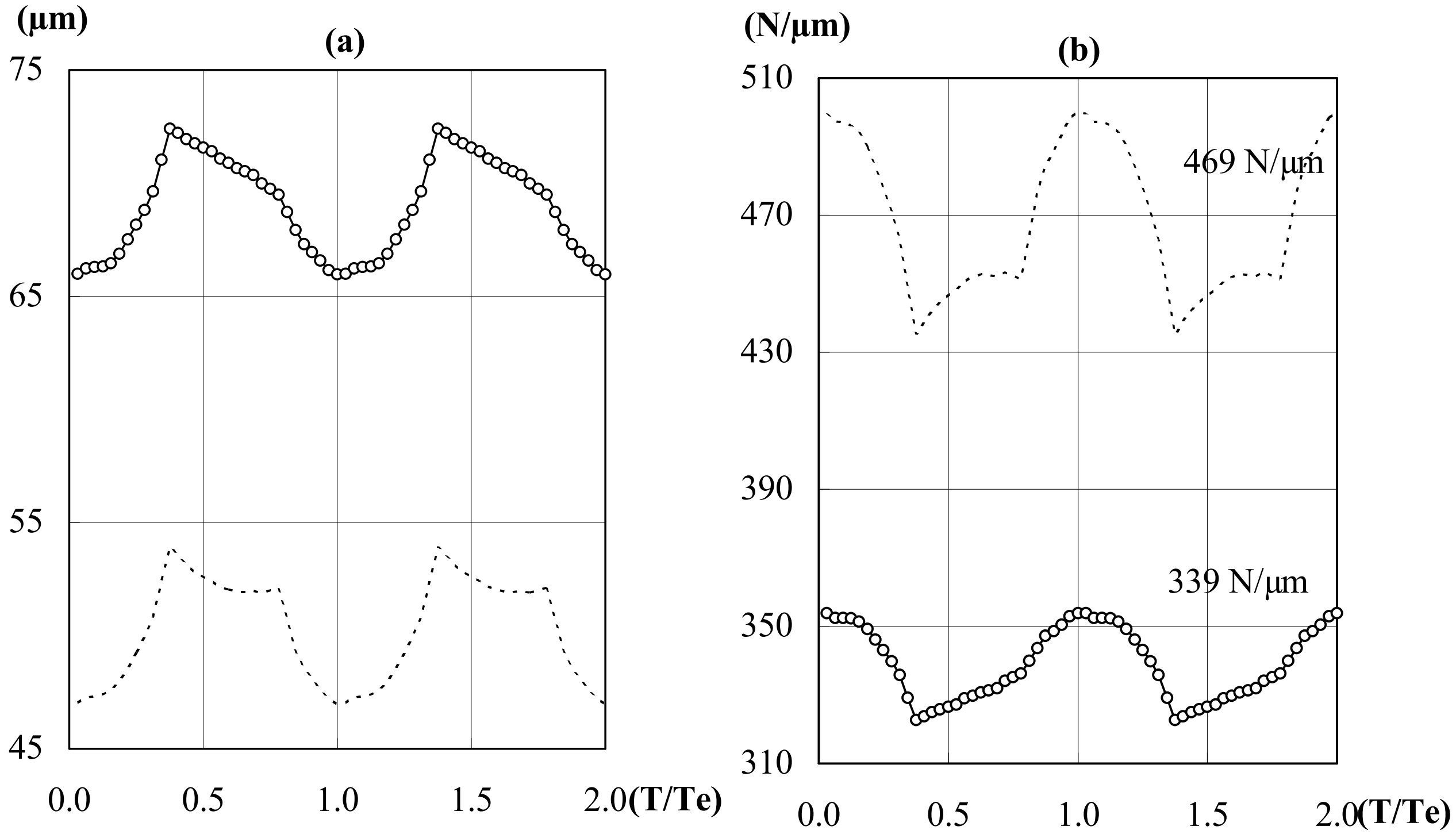

Figure 3. Evolutions temporelles de l'erreur statique de transmission sous charge **(a)** et de la raideur d'engrènement **(b)**.

Dents adjacentes couplées (oooooooo) et dents adjacentes non couplées (--------).

*Figure 3. Static transmission error* **(a)** *and mesh stiffness* **(b).**

*Interactions between adjacent loaded teeth on* (oooooooo) *and off* (--------).

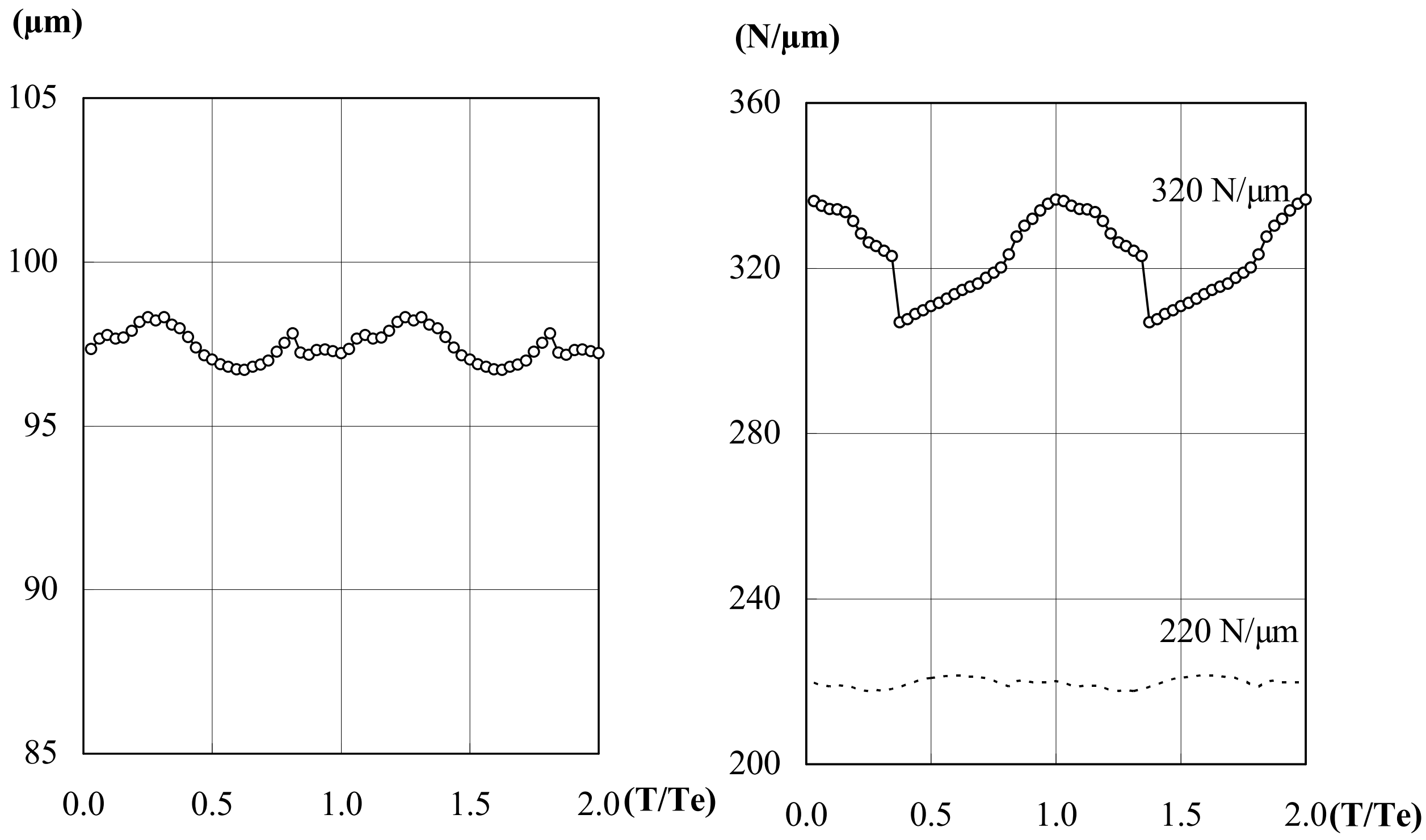

Figure 4. Evolutions temporelles de l'erreur statique de transmission sous charge **(a)** et de la raideur d'engrènement **(b)**.

$K(F_S, \theta) = \partial(F_S) / \partial(\delta(F_S, \theta))$ (oooooooo) et $K(F_S, \theta) = F_S / \delta(F_S, \theta)$ (--------).

*Figure 4. Static transmission error* **(a)** *and mesh stiffness* **(b).**

$K(F_S, \theta) = \partial(F_S) / \partial(\delta(F_S, \theta))$ (oooooooo) and $K(F_S, \theta) = F_S / \delta(F_S, \theta)$ (--------).